\documentclass[12pt,a4paper]{article}
\usepackage{graphicx}
\usepackage{mathtools}
\usepackage{xcolor}
\usepackage{amssymb}
\usepackage{mathrsfs}
\title{\Large\bf Arches of chaos, heteroclinic connections of first-order MMRs and the chaotic transport of small bodies in the Sun-Jupiter system}

\author{
Alessia Francesca Guido$^{1,2}$ and Christos Efthymiopoulos$^{3}$\\
\small{$^1$ Department of Physics, University of Trento}\\
\small{$^2$ Department of Mathematics, University of Rome Tor Vergata}\\
\small{$^3$ Department of Mathematics Tullio Levi-Civita, University of Padova}
}
\date{}
 
\begin{document}

\maketitle

\abstract{We investigate the heteroclinic connections between the stable and unstable manifolds of the unstable periodic orbits associated with the most important mean motion resonances (MMRs) in the Sun-Jupiter planar restricted three-body problem. In particular, we explicitly compute the stable and unstable manifolds of the unstable periodic orbits associated with the first order interior MMRs 2:1, 3:2, and the exterior MMR 2:3. Furthermore, we compute short-time FLI maps showing the chaotic saddle structure created by the stable and unstable manifolds of several interior or exterior MMRs other than the 1:1 (co-orbital) resonance. Transits of particles from the exterior to the interior of the orbit of Jupiter and vice versa are allowed for values of the Jacobi energy mapped to Tisserand parameter $T<3.0$. Such transits are shown to exist through a variety of heteroclinic channels. Besides the classical ones established by Koon et al. (\cite{koon2000},\cite{koonetal2001}), we give evidence of heteroclinic connections between the manifolds of the short-period orbits around L3 and of the periodic orbits associated with interior or exterior first order MMRs. Moreover, we observe direct heteroclinic connections between the manifolds of the interior with exterior MMRs, which do not involve the manifolds of any of the periodic orbits of the co-orbital resonance. Through the manifolds of the MMRs and the corresponding `ridges' in the numerical FLI maps, we explain the `arches-of-chaos' structures (\cite{todetal2020}) found in FLI maps in the asteroid plane of orbital elements $(a,e)$. The chaotic orbits shadowing the heteroclinic orbits exhibit `resonance hopping', indicating a possible connection to the behavior reported in literature as regards the orbits of several real Solar System objects classified as quasi-Hildas (QH) or Jupiter-family comets (JFC). Most of our results are obtained in the framework of the planar circular RTBP. However, through FLI maps we show that the manifold connections observed in the circular problem persist in the elliptic problem as well.}

\vspace{0.5em}
\section{Introduction}
\label{sec:intro}
The chaotic transport of small bodies from the neighborhood of one to that of another mean-motion resonance (MMR) is a well known phenomenon in Celestial Mechanics. Extended literature on this phenomenon can be found, steered by its key role in understanding the chaotic orbital behavior of asteroids or comet-like natural objects near MMRs in the Sun-Jupiter system (\cite{frosch1979},\cite{wis1980},\cite{dunqui1993},\cite{fraetal1993},\cite{fra1995},\cite{feretal1996},\cite{michfer1995},\cite{belmar1997},\cite{nesfer1997},\cite{roigetal2002},\cite{deletal2005},\cite{disisetal2005},\cite{baimal2009},\cite{schu2009},\cite{gargil2018},\cite{todetal2020},\cite{kazkaz2021},\cite{chaetal2022},\cite{dvokub2022},\cite{panhou2022},\cite{oldetal2023},\cite{coretal2024}, as well as the chaotic transitions through MMRs for the cislunar orbits of artificial bodies in the Earth-Moon system (\cite{beletal2008},\cite{top2008},\cite{gupkum2016},\cite{leixu2018},\cite{kumetal2024}). 

Present theories of the chaotic transport through MMRs lie in two basic pilars. 

First is the application of Chirikov's resonance overlap theory \cite{chi1979} to the MMRs in our solar system (\cite{wis1980}). This theory computes the size of the separatrix width associated with MMRs $q:p$ of any low order $r=|q-p|$. Wisdom's heuristic `$\mu^{-2/7}$' law applies to first order MMRs ($r=1$), and it has been used to estimate the relative width $\delta a/a_J$ in semi-major axis of the chaotic zone around Jupiter ($\mu=0.001$) formed by resonance overlap between the MMRs of first order embedded in the `outer asteroid belt', i.e., beyond the 2:1 MMR. Wisdom's law has been revised, both by analytical and numerical methods, to compare with numerical estimates as well as to include the dependence of the size of the separatrices on the small particles' orbital eccentricity or inclination (\cite{dun1989},\cite{quifab2006},\cite{muswya2012},\cite{decketal2013},\cite{rametal2015}, see also \cite{murder1999} for a general review). 

As regards numerical results on the overlapping of resonances in the outer asteroidal belt, several studies (see references above) establish the presence of a connected chaotic component transcending essentially the entire phase-space beyond the 2:1 resonance, connecting the 3:2 (Hildas) region with the co-orbital (1:1) resonance, as well as the co-orbital resonance with MMRs exterior to the orbit of Jupiter, i.e., the 2:3 and 1:2 resonances. The depletion of the MMR zones in the outer asteroid belt depends on the time required for chaotic transport to become effective, i.e., to allow particles to abandon the MMRs and enter into the direct (Hill) zone of planet close encounter. The empirical `Lecar's law' (\cite{lecetal1992},\cite{fraetal1993}) establishes a correlation between the particles' survival time and the mean Lyapunov time of most chaotic trajectories within the resonance overlap domain, except for those exhibiting `stable chaos' \cite{milnob1993}\cite{miletal1997}. When individual chaotic trajectories are considered, it is found that particles within the resonance overlap domain often undergo several transitions, or `jumps', from one MMR to the other. This phenomenon, called `resonance hopping' by \cite{belmar1997} (see also \cite{beletal2008},\cite{leixu2018},\cite{panhou2022}), is a typical behavior reported in literature also for classes of small objects categorized as `quasi-Hildas' (QH), 'Jupiter-family comets' (JFC) or `Centaurs' (\cite{disisetal2005},\cite{deletal2005},\cite{schu2009},\cite{baimal2009},\cite{gargil2018},\cite{todetal2020},\cite{kazkaz2021},\cite{dvokub2022},\cite{oldetal2023},\cite{coretal2024}). 

Our current understanding of phenomena as the above relies on a second pilar, namely the theory of heteroclinic transitions through the manifolds of the unstable Lyapunov orbits around the Lagrangian points L1 and L2 developed by Koon et al. \cite{koon2000},\cite{koonetal2001}. Denoting PL1 and PL2 the horizontal unstable Lyapunov periodic orbits around L1 and L2 respectively, \cite{koon2000} rigorously establishes that the stable manifolds of PL1 intersect with the unstable manifolds of PL2 (and vice versa) at the neighborhood of the interior MMR 3:2 as well as the neighborhood of the exterior MMR 2:3. These intersections are summarized in figure 6 of \cite{koonetal2001}. In physical terms, such heteroclinic connections imply that particles from the neighborhood of the 3:2, or 2:3, resonance can approach the neighborhood of L1 or L2, and vice versa. Through the shadowing lemma, this establishes existence of a heteroclinic chain connecting the involved MMRs through the transition $3:2\rightarrow 1:1 \rightarrow 2:3$. 

As mentioned in \cite{koonetal2001}, resonance connections similar to the above ``should exist for all nearby energies, as confirmed by numerical experiment''. In the present paper we give numerical evidence that, for small bodies in the Sun-Jupiter system, the above statement in fact applies to the heteroclinic connections associated with the manifolds not only of the PL1 and PL2 periodic orbits, but, practically, of the periodic orbits associated of any other low order MMR in the domain where these resonances overlap. Such heteroclinic connections then offer a natural framework to interpret the dynamical behavior of Solar System small bodies whose trajectories exhibit resonance hopping and/or transits through the orbit of Jupiter. Recently, heteroclinic connections similar to those here discussed were reported also in the Earth-Moon-particle problem (\cite{kumetal2024}), thus their appearance in the framework of the restricted three body problem appears rather generic, for sufficiently large values of the mass parameter $\mu$.    

Returning to the Sun-Jupiter case, taking as basic model the planar and circular restricted three-body problem, we may consider the totality of the stable and unstable manifolds emanating from the unstable periodic orbits of all MMRs $q:p$, with $q>p$ for interior resonances, $q<p$ for exterior resonances, and $q=p=1$ for the co-orbital resonance. In the latter case, a simple analysis of phase portraits in the entire domain of the co-orbital resonance (see section \ref{sec:fli} below) shows that the dominant chaotic zone around the islands of stability associated with Trojan asteroids is delimited by the lobes of the asymptotic manifolds of the PL3 unstable periodic orbit. The manifolds of the PL1 and PL2 orbits delimit, instead, the domain associated with Quasi-Satellite (QS) orbits. Consider a section $\mathcal{S}_{\ell_0,g_0}$ in phase space corresponding to fixed angles $\ell=\ell_0$ (mean anomaly), and $g=g_0$ (argument of the perihelion), as well as a fixed value of the Jacobi energy $E_J$. Denote by $\mathcal{W}^{S}_{q:p,E_J}$ and $\mathcal{W}^{U}_{q:p,E_J}$ the stable (unstable) tube manifolds of the unstable periodic orbit corresponding to the $q:p$ resonance, if $q\neq p$, or to the PL3 orbit, if $q=p=1$. The manifolds $\mathcal{W}^{S}_{q:p,E_J}$ are two-dimensional (`tubes'), hence they intersect the section $\mathcal{S}_{\ell_0,g_0}$ at curves $\mathcal{C}^{S}_{q:p,E_J}$ labeled by the value of the Jacobi energy. Such curves can then be projected onto the plane of the remaining (non-fixed) Keplerian elements $(a,e)$ (semi-major axis, eccentricity). As discussed in \cite{guzleg2014} (see also the review \cite{guzleg2023} and references therein, as well as section \ref{sec:fli} below), such intersections of the manifolds with fixed suitably defined section surfaces are in principle possible to visualize without explicit computation of the manifolds $\mathcal{W}^{S}_{q:p,E_J}$, using short-time Fast Lyapunov Indicator (FLI) maps. In particular, consider the family of all curves $\mathcal{C}^{S}_{q:p,E_J}$ intersecting a given parallelogram $a_{min}\leq a\leq a_{max}$, $e_{min}\leq e\leq e_{max}$, with $E_J$ in the permissible range consistent with the above parallelogram, and $(q,p)$ any positive integers satisfying some maximum order $|q-p|\leq Q_{max}$ of the considered MMRs (say $Q_{max}=2$). Then, a numerical computation of short-time FLI maps for trajectories with initial conditions in the above domain and integrated forward should visualise simultaneously the entire above family of curves $\mathcal{C}^{S}_{q:p,E_J}$. Similarly, a visualization of the entire family of curves $\mathcal{C}^{U}_{q:p,E_J}$, corresponding to the intersections of the unstable tube manifolds $\mathcal{W}^{U}_{q:p,E_J}$ with the surface of section  $\mathcal{S}_{\ell_0,g_0}$ should be possible by computing the FLI maps of trajectories in the domain integrated backward in time. 

Stability maps for forward-integrated trajectories, using the short-time FLI or other indicators of chaos, have been presented in the context of studies of the orbital stability or the chaotic transport of small bodies in the framework of the Sun-Jupiter or Earth-Moon RTBP (\cite{rametal2015},\cite{todetal2020},\cite{cavallariefhty},\cite{rossiefthy}; see also figure 2.31 and associated discussion in \cite{FerrazMello2024}). A comprehensive view of the entire manifold structure obtained through such maps is shown in figure 1 of \cite{todetal2020}, called by them the `arches of chaos'. Most notably, visual comparison of the bottom and top panels of figure 1 of \cite{todetal2020} shows that the manifold structure obtained through integration of the trajectories in the circular RTBP remains practically unaltered when passing to more precise models of the equations of motion including the direct and indirect effects to asteroidal trajectories by all the giant planets of the Solar system. Note also that, as established in \cite{gawmar2009}, the total manifold structure seen in such figures, computed through maps based on chaotic indicators, is the analog, for a Hamiltonian phase-space flow, of the `Lagrangian Coherent Structures (LCS)' found in fluids described through nonlinear vector fields (\cite{hal2002}).  

In the present paper, we aim to specify the exact correspondence between the structures observed in the numerical (e.g. FLI) stability maps and the manifolds $\mathcal{W}^{S}_{q:p,E_J}$ and $\mathcal{W}^{U}_{q:p,E_J}$, through a latter's \textit{explicit computation} by common numerical techniques of manifold computation in dynamical systems. To this end, we choose different values of the Jacobi energy $E_J$, as well as a `pericentric' Poincar\'{e} map (\cite{ursgal2013},\cite{paeeft2015},\cite{wanmal2017}),  which can be mapped one-to-one to the surface of section used in \cite{koon2000}. Then, we first explicitly compute the fixed points of the unstable periodic orbits $q:p$ for the MMRs $2:1$, $3:2$ (interior), PL3 (co-orbital) and $2:3$ exterior in the previous Poincar\'{e} map, as well as the stable and unstable manifolds $\mathcal{W}^{S}_{q:p,E_J}$ and $\mathcal{W}^{U}_{q:p,E_J}$ emanating from each of the previous orbits. Finally, we compute numerical short-time FLI maps of trajectories integrated forward or backward in time, and with initial conditions in the same surface of section, and discuss how the structure seen in the FLI maps compares with the superposition of all explicitly computed manifolds, $\mathcal{W}^{S}_{q:p,E_J}$ or $\mathcal{W}^{U}_{q:p,E_J}$. 

Besides establishing the correspondence between the `arches of chaos' and the manifolds $\mathcal{W}^{S}_{q:p,E_J}$ and $\mathcal{W}^{U}_{q:p,E_J}$, our study leads to several results pertinent to the way we interpret the chaotic transport of small particles through the orbit of Jupiter: (i) we find that for values of $E_J$ corresponding to Tisserand parameters $2\leq T_J\leq 3$, the low-order lobes of the manifolds of the $PL3$ family penetrate deep inside the zones of inner MMRs up to 2:1, and outer up to 2:3. In terms of Keplerian elements, the corresponding periodic orbits $P_{q:p}$ have eccentricities large enough so that they reach (to within one to few Hill spheres) the distance of the co-orbital resonance at apocenter, for interior MMRs, or pericenter, for exterior MMRs. The trajectories shadowing these manifolds are chaotic. However, analysis of the phase portraits suggests that the source of chaos in this case should be properly attributed to the overlapping of the MMRs rather than to close encounters with Jupiter. The separatrix-like chaotic zones of the co-orbital resonance formed by the manifolds of the Lyapunov orbits PL3 make a substantial part of the above resonance overlap domain. Also, the usually considered manifolds of the Lyapunov orbits PL1 and PL2 are involved in the type of chaotic orbits here discussed mostly through their heteroclinic intersections with the manifolds of the PL3 family. On the other hand, the manifolds of the families PL1 and PL2 are also (and principally) involved to direct close encounters of the small bodies with Jupiter, which, besides transits through the Hill domain, may also lead to temporary captures around Jupiter, as discussed in \cite{tanetal1990} (see also \cite{todetal2020}). (ii) We find direct heteroclinic intersections between the manifolds of the innermost and the outermost of our considered first order MMRs, i.e., 2:1 and 2:3 respectively. Such intersections can be interpreted by the fact that the entire family of manifolds $\mathcal{W}^{S}_{q:p,E_J}$ (or $\mathcal{W}^{U}_{q:p,E_J}$) for different energies and different periodic orbits forms a unique Lagrangian Coherent Structure, in which direct heteroclinic paths between an inner and an outer MMR, as for example $2:1\rightarrow 2:3$ move parallelly to the indirect ones involving the co-orbital resonance, such as $2:1\rightarrow  L3\rightarrow 2:3$. (iii) We compute forward and backward FLI maps in the elliptic RTBP, giving numerical indications that the manifold structures explicitly computed in the circular problem survive essentially unaltered in the elliptic one. This verification is useful, since an explicit computation of the manifolds in the elliptic or higher-complexity problems requires prior explicit computation of the associated low-dimensional partially hyperbolic invariant tori, which generalize the periodic orbits $P_{q:p}$ of the circular problem. Examples of such computation in particular cases, as for example, for the orbits around the points L1 and L2 of the elliptic RTBP are given in \cite{jornic2020},\cite{fitros2022},\cite{paeguz2023}. However, a generic computation of the kind for the family of low-dimensional tori generalizing the periodic orbits of the whole set of first-order MMRs appears rather hardeous to repeat in any problem surpassing the complexity of the planar and circular RTBP.  

The paper is structured as follows: section \ref{sec:model} gives all basic definitions whose use is made in the rest of the paper. Section \ref{sec:fli} shows numerical phase portraits and the corresponding short-time FLI maps, allowing to numerically visualizing the underlying structure created by the manifolds of the most important interior and exterior MMRs. Section \ref{sec:manifolds} shows the explicit computation of the manifolds of the unstable periodic orbits $P_{q:p}$, and makes an extended discussion on the interpretation of their formed heteroclinic connections, direct, i.e., not involving, or indirect, i.e., involing the manifolds of the periodic orbits of the co-orbital resonance. Section \ref{sec:ertbp} shows how the manifold structure found through the FLI maps generalizes when passing from the circular to the elliptic RTBP. Section \ref{sec:conclusions} summarizes the basic conclusions from the present study. 

\section{Model}
\label{sec:model}

\subsection{Circular RTBP}
\label{subsec:crtbp}
Our adopted basic model is the Sun-Jupiter planar circular restricted three-body problem (CRTBP). The Hamiltonian of particle motion in cylindrical coordinates $(r,\varphi)$ in the barycentric synodic frame is
\begin{align}\label{eq:hamcirc}
H &= \frac{1}{2} (p_{r}^2+\frac{p_{\varphi}^2}{r^2})-\Omega_J p_{\varphi} \\
  &- \frac{\mathcal{G}M_{S}}{\sqrt{(r\cos\varphi-x_{S})^2+(r\sin\varphi-y_{S})^2}}-\frac{\mathcal{G}M_{J}}{\sqrt{(r\cos\varphi-x_{J})^2+(r\sin\varphi-y_{J})^2}}~~,\nonumber
\end{align}
where $x_S=-a_JM_J/(M_S+M_J)$, $y_S=0$, $x_J=a_JM_S/(M_S+M_J)$, $y_J=0$, $M_S=1M_\odot$ (mass of the Sun), $M_J=0.00096 M_\odot$ (mass of Jupiter), 
$a_{J}=5.19~AU$ (heliocentric semi-major axis of Jupiter), $\mathcal{G}=4\pi^2 AU^3/(yr^2M_\odot)$ (Newton's constant) and $\Omega_J=\left(\mathcal{G}(M_S+M_J)/a_J^3\right)^{1/2}$ (Jupiter's mean motion). We use cylindrical coordinates, since they are convenient in for defining a suitable Poincar\'{e} surface of section useful in illustrating phase portraits as well as in the computation of the periodic orbits associated with each examined mean motion resonance (see below). In the Hamiltonian (\ref{eq:hamcirc}), the canonical momenta $p_r=\dot{r}$, $p_\varphi=r^2(\dot{\varphi}+\Omega_J)$ represent the radial velocity and specific angular momentum of a particle in a \textit{rest} frame of reference whose axes instantaneously coincide with the axes of the barycentric synodic reference frame. 

The numerical value of the Hamiltonian (\ref{eq:hamcirc}) is a constant of motion (Jacobi energy). Far from Jupiter, the Jacobi energy is related to the particle's Tisserand parameter
\begin{equation}\label{eq:tisserand}
T_J={a_J\over a}+2\sqrt{\left({a\over a_J}\right)(1-e^2)}
\end{equation}
through the relation
\begin{equation}\label{eq:tissjac}
T_J=-2C_J
\end{equation}
where
\begin{equation}\label{eq:jacobi}
C_J =\left({a_J\over\mathcal{G}(M_S+M_J)}\right) H(r(t), \varphi(t), p_r(t), p_\varphi(t))=\left({a_J\over\mathcal{G}(M_S+M_J)}\right)  E_J
\end{equation}
is the Jacobi constant, equivalent to the rescaled Jacobi energy $E_J=H$.  

We integrate Hamilton's equations of motion under the Hamiltonian (\ref{eq:hamcirc}), together with the associated variational equations, using a standard fourth-order ODE integrator with adjustable timestep in Matlab, with control of the precision up to about ten significant figures. Through numerical integration we then compute numerical short time Fast Lyapunov Indicator (FLI) maps. For the latter we use the definition (\cite{froschetal2000}):
\begin{equation}\label{eq:fli}
\mbox{FLI}(T)=\max_{0\leq t\leq T}\log_{10}\|\|\xi(t)\|\|
\end{equation}
where $\xi(t)$ is the length of the deviation vector at time $t$, starting from initial conditions of the deviation vector such that $\xi(0)=1$, and $T$ is the total integration time. In \cite{guzleg2014} it was suggested that a computation of the FLI for short times $T$ over a grid of initial conditions in phase space, collecting along each individual orbit only those parts of the growth of the deviation vector when the orbit comes close to a known (partially) hyperbolic low-dimensional invariant object (e.g., a computed periodic orbit) allows to visualize in phase space the invariant manifolds of that particular object. In particular, computing, as above,  FLI maps (value of the FLI against value of the initial conditions) in which the orbits are \textit{forward} integrated allows to visualize the \textit{stable} manifolds of the object in study, while the FLI maps of \textit{backward}-integrated orbits visualize the \textit{unstable} manifolds of the object in study. The collection of the relevant parts of the orbits is done through a window function which imposes a supression of the growth of the deviation vector outside an open domain in phase space around the invariant object of interest. Since in our case we are interested in visualizing the manifolds (stable or unstable) of many different unstable periodic orbits, associated with different MMRs, here we introduce no window function, but simply control the integration time $T$ to be small enough, i.e., of few decades of periods of the periodic orbits of interest. This is achieved in practice by setting $T=100~$yrs.

\subsection{Elliptic RTBP}
\label{subsec:ertbp}
In section \ref{sec:ertbp} we show some short-time FLI maps computed in the planar elliptic RTBP. These are computed by forward or backward integration of trajectories in the same equations of motion (and variational equations) as those under the Hamiltonian (\ref{eq:hamcirc}), but with the replacement $x_J\rightarrow x_J(t)$, $y_J\rightarrow y_J(t)$, $x_S\rightarrow x_S(t)$, $y_S\rightarrow y_S(t)$, where 
\begin{align}\label{eq:rotelliptic}
x_J(t)&=~~X_J(t)\cos(\Omega_J t)+Y_J(t)\sin(\Omega_J t) \nonumber\\
y_J(t)&=-X_J(t)\sin(\Omega_J t)+Y_J(t)\cos(\Omega_J t) \\
x_S(t)&=~~X_S(t)\cos(\Omega_J t)+Y_S(t)\sin(\Omega_J t) \nonumber\\
y_S(t)&=-X_S(t)\sin(\Omega_J t)+Y_S(t)\cos(\Omega_J t) \nonumber
\end{align}
where the inertial frame coordinates $\vec{R}_J=(X_J,Y_J)$, $\vec{R}_S$ are derived by the evolution of Jupiter's heliocentric position vector $\vec{R}(t)$ according to: $\vec{R}_J=M_S\vec{R}/(M_S+M_J)$, $\vec{R}_S=-M_J\vec{R}/(M_S+M_J)$, and $\vec{R}(t)$ satisfying the differential equation $\ddot{\vec{R}}(t)=-\mathcal{G}(M_S+M_J)\vec{R}/R^3$. We solve the latter using the initial condition $\vec{R}(0)=(a_J(1-e_J),0)$, $\dot{\vec{R}}(0)=(0,\sqrt{\mathcal{G}(M_S+M_J)(1+e_J)/(a_J(1-e_J))})$, with $e_J=0.0487$, i.e., launching the system from pericentric position of the two bodies at the time $t=0$. We obtain the solution directly through a numerical solution of Kepler's equation. Physically, we keep viewing the motion of the Sun and Jupiter in a baricentric frame which rotates uniformly, with angular speed equal to the mean motion $\Omega_J$. In this rotating frame both Jupiter and the Sun describe small retrograde epicycles which have an elliptic form, with major axis equal to $2a_J'e_J$, $2a_Se_J$ and minor axes equal to $a_J'e_J$ and $a_S'e_S$ respectively, where $a_S'=M_Ja_J/(M_S+M_J)$, $a_J'=M_Sa_J/(M_S+M_J)$. For the purposes of the present paper, this choice is preferable over integration in the rotopulsating frame of reference, since it allows to compute the FLI maps using precisely the same grid of initial conditions as in the circular case (see section \ref{sec:ertbp}).

\section{Phase portraits and FLI maps}
\label{sec:fli}
\subsection{Poincar\'{e} map}
\label{subsec:poincare}
Explicit computations of phase portraits, as well as of the periodic orbits $P_{q:p}$ and their manifolds, require introducing a convenient Poincar\'{e} map intersected iteratively by all orbits of interest. Our choice is to map the orbits of fixed Jacobi energy $E_J$ in a pericentric surface of section $\mathcal{P}_{E_J}$ (\cite{ursgal2013},\cite{paeeft2015},\cite{wanmal2017}) defined by the relations 
\begin{align}\label{eq:poinc}
\mathcal{P}_{E_J}=
\bigg\{&p_r=0,~(\varphi,p_\varphi)\in(\mathbb{T}\times\mathcal{D}_{E_J}\subseteq\mathbb{R}^+),\\
~&r=\mbox{root of $H(\varphi,r,p_{r}=0,p_\varphi)=E_J$ 
with $\dot{p}_r=-{\partial H(\varphi,r,p_\varphi,p_r=0)\over\partial r}>0$}\bigg\}~. \nonumber
\end{align}
The subset $\mathcal{D}_{E_J}$ is determined by the requirements that the orbit should be prograde ($p_\varphi>0$) and that there exists a root for $r$ satisfying the conditions of (\ref{eq:poinc}). A Poincar\'{e} map can be defined in the above section:
\begin{equation}\label{eq:poincmap}
\mathcal{M}_{E_J}:~
(\mathbb{T}\times\mathcal{D}_{E_J})
\rightarrow
(\mathbb{T}\times\mathcal{D}_{E_J})
\end{equation}
by choosing a pair of values $(\varphi,p_\varphi)\in(\mathbb{T}\times\mathcal{D}_{E_J})$, computing the corresponding point in phase space belonging to the section $\mathcal{P}_{E_J}$ (through Eq.(\ref{eq:poinc})), integrating the orbit numerically until it intersects again the section $\mathcal{P}_{E_J}$, and collecting the values of the pair $(\varphi',p_\varphi')$ at the new intersection. This effectively defines the symplectic map $\mathcal{M}_{E_J}$ through numerically evaluated functions $F_{\varphi}^{E_J}: (\mathbb{T}\times\mathcal{D}_{E_J})\rightarrow\mathbb{T}$, $F_{p_\varphi}^{E_J}: (\mathbb{T}\times\mathcal{D}_{E_J})\rightarrow\mathcal{D}_{E_J}$, such that 
\begin{equation}\label{eq:poincmap2}
\varphi'=F_{\varphi}^{E_J}(\varphi,p_\varphi),~
p_\varphi'=F_{p_\varphi}^{E_J}(\varphi,p_\varphi)~~.
\end{equation} 
Through the functions $(F_{\varphi}^{E_J},F_{p_\varphi}^{E_J})$ we may further explicit compute the periodic orbits $P_{q:p}$ as fixed points of the Poincar\'{e} map (\ref{eq:poincmap2}), and their manifolds by linearization around each fixed point (see section \ref{sec:manifolds}). Note that at any point in the section we have the relations $p_\varphi=G\sqrt{\mathcal{G}a(1-e^2)}$, while $\varphi=\overline{g}=g-\lambda_J$ (orbit at pericenter), where $(a,e,g)$ are the particle's osculating semi-major axis, eccentricity and argument of the perihelion in a fixed (inertial) frame coinciding with the rotating barycentric frame at the time $t=0$, $(G,g)$ are the corresponding Delaunay action angle variables, and $\lambda_J$ the mean longitude of Jupiter in the same frame. Hence, the above Poincar\'{e} map is similar to the one used in \cite{koon2000}, with $G$ used instead of $L$.

As regards our choice of values of the Jacobi energy $E_J$, this is representative of the distribution of Tisserand parameters reported in literature for QHs or JFCs. According to \cite{toth2006}, the bulk of quasi-Hildas are in Tisserand parameters ranging from $T_J=2.9$ to $T_J=3.05$. The distribution of Tisserand parameters of evolved JFCs with initial heliocentric pericentric distance $q<a_J=5.2$~AU is somewhat broader, i.e. $2.8<T_J<3.1$, while, from \cite{ricketal2017}, the distributions of $T_J$ of the objects in a simulated scattered disc evolved by the `Nice model' is still broader, $2.4<T_J<3$. Here we focus on the interval of Jacobi constants $-1.4<C_J<-1.525$, which corresponds to Tisserand parameters $2.8<T_J<3.05$.  Of course, we emphasize that our study is limited to the planar RTBP, while several of the real objects mentioned above have important inclinations. Hence, albeit probable, the connection between our results and the orbital behavior of observed real Solar System small bodies in the above classes should be regarded with due caution.  

\begin{figure}[h]
\centering
\includegraphics[scale=0.7]{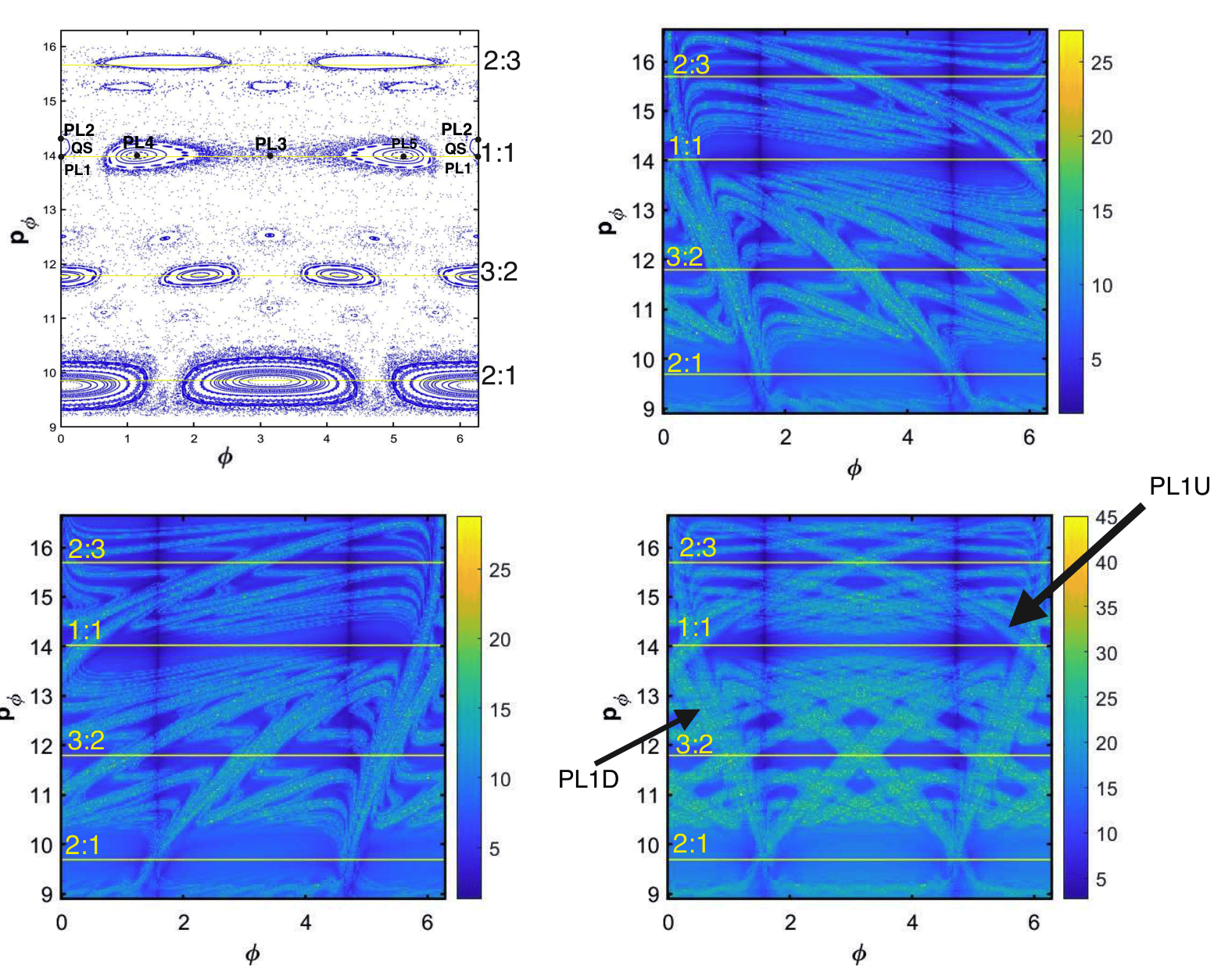}
\caption{\small Comparison of the phase portrait and the manifold structure for the Jacobi energy $E_J=-1.48$, corresponding to the Tisserand parameter $T_J = 2.96$. \textit{Top  left:} The phase portrait in the Poincar\'e section $\mathcal{P}_{E_J=-1.48}$ obtained by numerical integration of several trajectories with initial conditions in the square $\varphi \in [0,2\pi]$ and $p_\varphi \in [9,16]$. \textit{Top right:} color-scaled FLI map for the backward-integrated trajectories with initial conditions in a $300\times 300$ grid in the above square and integration time down to $t=-100~yrs$. \textit{Bottom left:} FLI map (same as above) but for forward-integrated trajectories up to the time $t=100~yrs$. \textit{Bottom right:} Superposition of the  FLI maps for the forward and backward-integrated trajectories.}
\label{fig:portraitfli148}
\end{figure}

Figure \ref{fig:portraitfli148} top-left shows an example of phase portrait by the Poincar\'{e} map computed as above for Jacobi constant $C_J=-1.48$. This value is slightly above the one of the Lagrangian points $L1,L2$, hence communication of the domains interior and exterior to the orbit of Jupiter is energetically allowed. The figure shows the islands of stability around the interior first order MMRs 2:1, 3:2, 4:3, the exterior first order MMRs 2:3 and 1:2, as well as the second order MMRs 5:3 (interior) and 3:5 (exterior). The entire phase-space structure of the co-orbital resonance is also distinguished. The main (trojan) islands of stability are around the short-period stable orbits PL4,PL5, separated by the separatrices of the short-period unstable periodic orbit PL3. On the other hand, the separatrices of the Lyapunov orbits PL1 and PL2 delimit a second smaller island of stability around the stable periodic quasi-satellite (QS) orbit. 

The remaining panels in this figure illustrate, now, comprehensively the rich heteroclinic network formed by the manifold structures embedded in phase space at this level of Jacobi energies. The manifold structures are revealed through FLI color maps computed by integrating the trajectories in a $300\times 300$ rectangular grid of initial conditions taken in the limits for $\varphi$ and $p_\varphi$ same as in the panel for a total integration time $T=100~$yrs. The top-right and bottom-left panels show the FLI maps computed with a backward- or forward-in-time integration of the orbits respectively. The lagrangian coherent structures formed by the union of the stable and unstable manifolds of several MMRs, both interior and exterior to the orbit of Jupiter, are visually straightforward to recognize in these figures. Furthermore, superposing the backward and forward FLI maps reveals a rich network of heteroclinic connections created by the intersections of these manifolds. Most notable are the classical heteroclinic paths rigorously studied in \cite{koon2000},\cite{koonetal2001}, formed by manifold structures departing from the neighborhood of the fixed points PL1 and directed downwards (marked PL1D in the figure) reaching the 2:1 resonance through the 3:2 resonance, or upwards (PL1U), reaching the 2:3 resonance, as well as the manifold structures departing from the neighborhood of the fixed point PL2 and directed upwards (PL2U), reaching the 2:3 resonance. The heteroclinic orbits in such paths are shadowed by orbits representing a `temporary satellite capture' (TSC, see \cite{tanetal1990}), and we point out their possible connection to the `Centaurs-JFC gateway' \cite{saridetal2019},\cite{todetal2020}).  

\begin{figure}[h]
\centering
\includegraphics[scale=0.7]{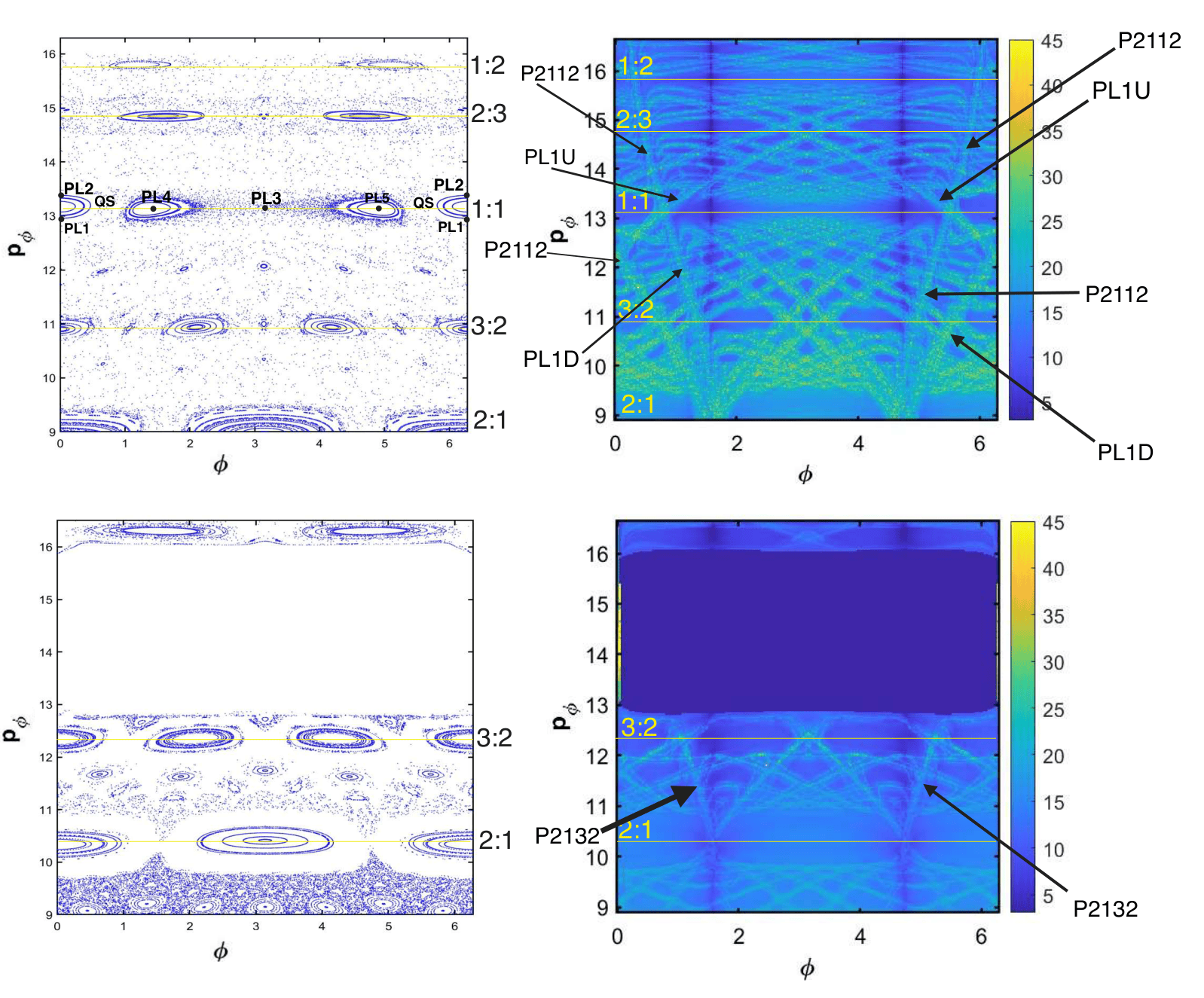}
\caption{\small Poincar\'e sections (left) and superposition of the FLI maps for forward and backward-integrated trajectories (right) at the Jacobi energies $E_J=-1.42$ (corresponding to $T_J=2.84$, top), or $E_J=-1.52$ (corresponding to $T_J=3.04$, bottom).}
\label{fig:portraitfli142152}
\end{figure}
Figure \ref{fig:portraitfli142152}, now, shows the phase portraits and superposed forward and backward short-time FLI maps for two values of the Jacobi constant near the edges of our considered range, i.e., $C_J=-1.42$ (corresponding to $T_J=2.84$, top row) and $C_J=-1.52$ ($T_J=3.04$, bottom row). 

For values of $T_J$ near the lower edge of the distribution, chaos overall increases and the size of the islands corresponding to stable motions at the MMRs decreases considerably. Comparing manifold structures and their heteroclinic connections, the most important difference regards the clear appearance, for smaller $T_J$ of a heteroclinic path denoted as P2112 in the figure, which appears to originate at the inner 2:1 MMR, and transcends the entire phase space, reaching the outer 2:3, and even 1:2 MMRs. Note that this path avoids the neighborhood of the PL1 and PL2 orbits. Note that this path seems to exist also for $T_J=2.96$ (cf. second panel of Fig.\ref{fig:portraitfli142152} with last panel of Fig.\ref{fig:portraitfli148}), but with a smaller distance from the path PL1D. In fact, the distance between the two paths is of the order of the horizontal size of the islands of quasi-satellite orbits, which, in turn, increases as the value of $T_J$ decreases. 

On the other hand, for values of $T_J$ near the higher edge of the considered distribution (e.g. $T_J=3.04$, bottom row of Fig.\ref{fig:portraitfli142152}), the domains interior and exterior to the orbit of Jupiter are energetically separated, and there can be no heteroclinic connection of any of the manifolds of an interior MMR with the manifolds of exterior MMRs. In this case, the only prominent heteroclinic connections observed through FLI maps are among the manifolds of the most important interior MMRs, as, for example, the path P2132 connecting the 2:1 and 3:2 MMRs. 

\section{Invariant manifolds and hetetoclinic connections of the first order MMRs}
\label{sec:manifolds}

\subsection{Computation of the manifolds associated to unstable periodic orbits}
In this section we superpose the structures seen in the FLI maps obtained in the previous section and the explicitly computed stable and unstable manifolds of the unstable periodic orbits associated with some major mean motion resonances of the first order, namely the inner resonances 2:1, 3:2, and the outer resonance 2:3. Also, we show in comparison the manifolds associated with the Lyapunov periodic orbit PL3 which emanates from the collinear point L3 for Jacobi energy values $E_J>E_{J,L3}$. 

As mentioned in the introduction, the stable and unstable tube manifolds $\mathcal{W}^{S}_{q:p,E_J}$, $\mathcal{W}^{U}_{q:p,E_J}$ of the unstable periodic orbit associated with the MMR $q:p$ intersect the Poincar\'{e} surface of section $\mathcal{P}_{E_J}$ at the curves $\mathcal{C}^{S}_{q:p,E_J}$, $\mathcal{C}^{U}_{q:p,E_J}$ respectively. Any point of intersection of the curves ${C}^{S}_{q:p,E_J}$, $\mathcal{C}^{U}_{q:p,E_J}$ defines the initial condition of a homoclinic orbit in the Poincar\'{e} section $\mathcal{P}_{E_J}$. Consider also two different MMRs $q:p$, $q':p'$. Any intersection of the curve $\mathcal{C}^{S}_{q:p,E_J}$ with $\mathcal{C}^{U}_{q':p',E_J}$, or of the curve $\mathcal{C}^{S}_{q':p',E_J}$ with $\mathcal{C}^{U}_{q:p,E_J}$, defines the initial condition of an heteroclinic orbit in the section $\mathcal{P}_{E_J}$, which tends asymptotically to one of the unstable periodic orbits $q:p$ or $q':p'$ in the asymptotic limit of infinity in the forward or backward sense of time.  

To approximate numerically the curves $\mathcal{C}^{S}_{q:p,E_J}$, $\mathcal{C}^{U}_{q:p,E_J}$, we first note that the unstable periodic orbit of the resonance $q:p$ has multiplicity $m=q$ in the surface of section $\mathcal{P}_{E_J}$, hence it corresponds to a fixed point of the $q-th$ iterate of the Poincar\'{e} return map (\ref{eq:poincmap2}). Using the Newton-Raphson method, we first compute this fixed point as well as the eigenvalues and eigenvectors of the monodromy matrix, i.e., the Jacobian matrix of the linearized $q-$iterared return map. Since we have no explicit formula but only a numerical evaluation of the functions $F_{\varphi}^{E_J}$, $F_{p_\varphi}^{E_J}$ of the return map (\ref{eq:poincmap2}), we use central finite differences to evaluate the Jacobian matrix $J(\varphi,p_{\varphi})=\partial(\varphi',p_{\varphi}')/\partial(\varphi,p_{\varphi})$ at any point of the surface of section, and check that the matrix $J(\varphi,p_{\varphi})$ satisfies the symplecticity condition to within eight significant figures. Then, the monodromy matrix of the unstable fixed point  $q:p$ is given by the product $\mathcal{A}_{q:p}=\prod_{i=1}^q J(\varphi_{q:p}^{(i)},p_{\varphi,q:p}^{(i)})$, where $(\varphi_{q:p}^{(i)},p_{\varphi,q:p}^{(i)})$, $i=1,\ldots q$ is the i-th iterate of the fixed point corresponding to the periodic orbit under the return map (\ref{eq:poincmap2}). Since the periodic orbit $q:p$ is unstable, the eigenvalues of the monodromy matrix $\mathcal{A}_{q:p}$ are real and satisfy the symplecticity condition $\lambda_1\lambda_2=1$. Taking initial conditions along a small segment (of length $\Delta S=10^{-3}$ along the unstable eigendirection starting from the fixed point, and iterating these initial conditions with the forward return map, visualizes the intersection of the unstable invariant manifold (curve $\mathcal{C}^{U}_{q:p,E_J}$) with the surface of section. Repeating the process with initial conditions along the stable eigendirection mapped with the backward return map visualizes the stable manifold (curve $\mathcal{C}^{S}_{q:p,E_J}$).

\begin{figure}[h]
\centering
\includegraphics[scale=0.7]{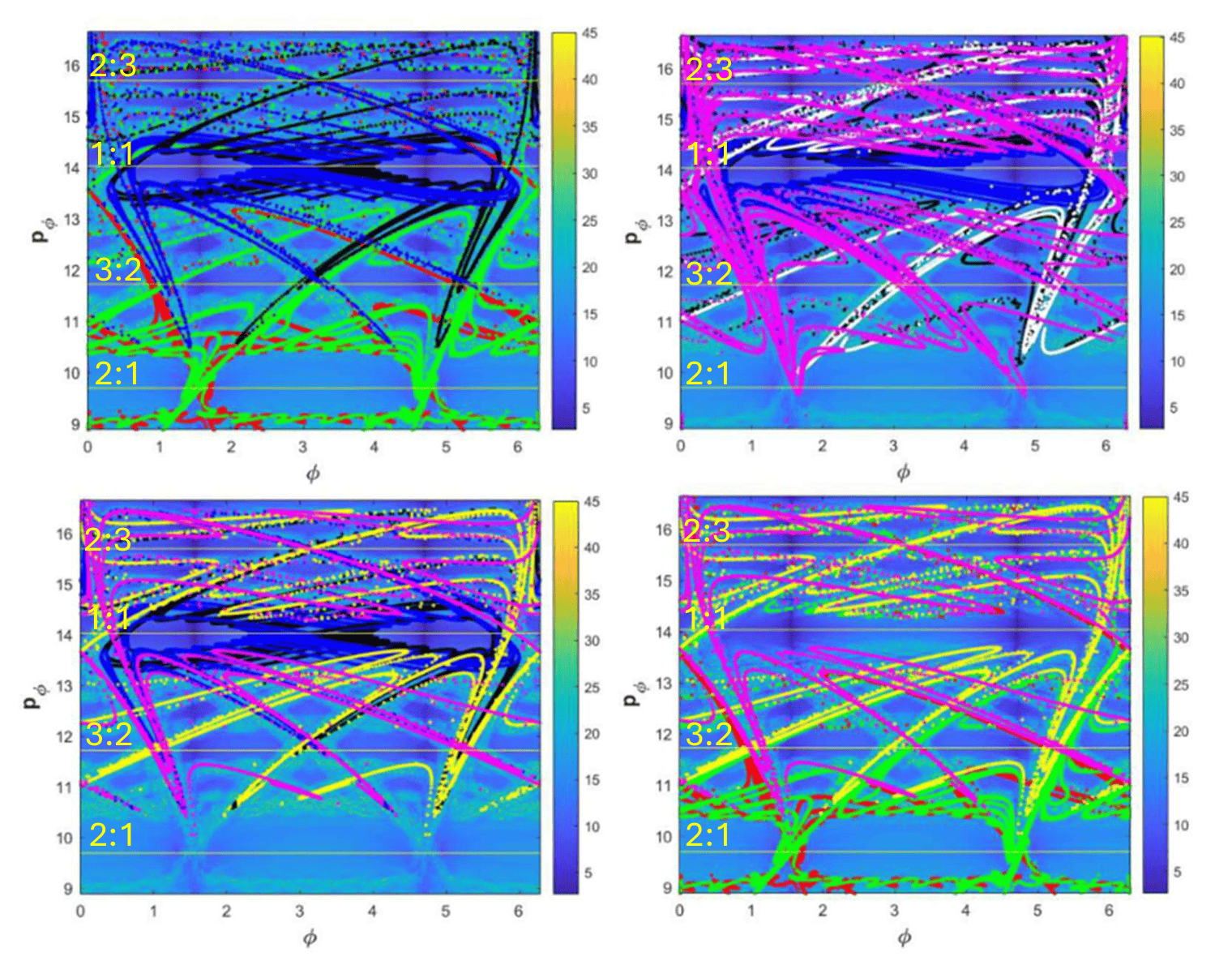}
\caption{\small Superposition of the (backward/forward integrated) FLI maps with the unstable/stable manifolds of the unstable periodic orbits of various MMRs $q:p$ at the Jacobi energy $E_J = -1.48$ ($T_J=2.96$). \textit{Top left:} Unstable/stable manifolds of the 2:1 resonance (red/green) superposed to those of the periodic orbit PL3 of the 1:1 resonance (blue/black). \textit{Top right:} Manifolds of the orbit PL3 of the 1:1 resonance (blue/black) superposed to those of the 3:2 resonance (magenta/white). \textit{Bottom left:} Manifolds of the orbit PL3 of the 1:1 resonance (blue/black) superposed to those of the 2:3 resonance (magenta/yellow). \textit{Top right:} Manifolds of the 2:1 resonance (red/green) superposed to those of the 2:3 resonance (magenta/yellow).}
\label{fig:flimanif}
\end{figure}
\begin{figure}[h]
\centering
\includegraphics[scale=0.5]{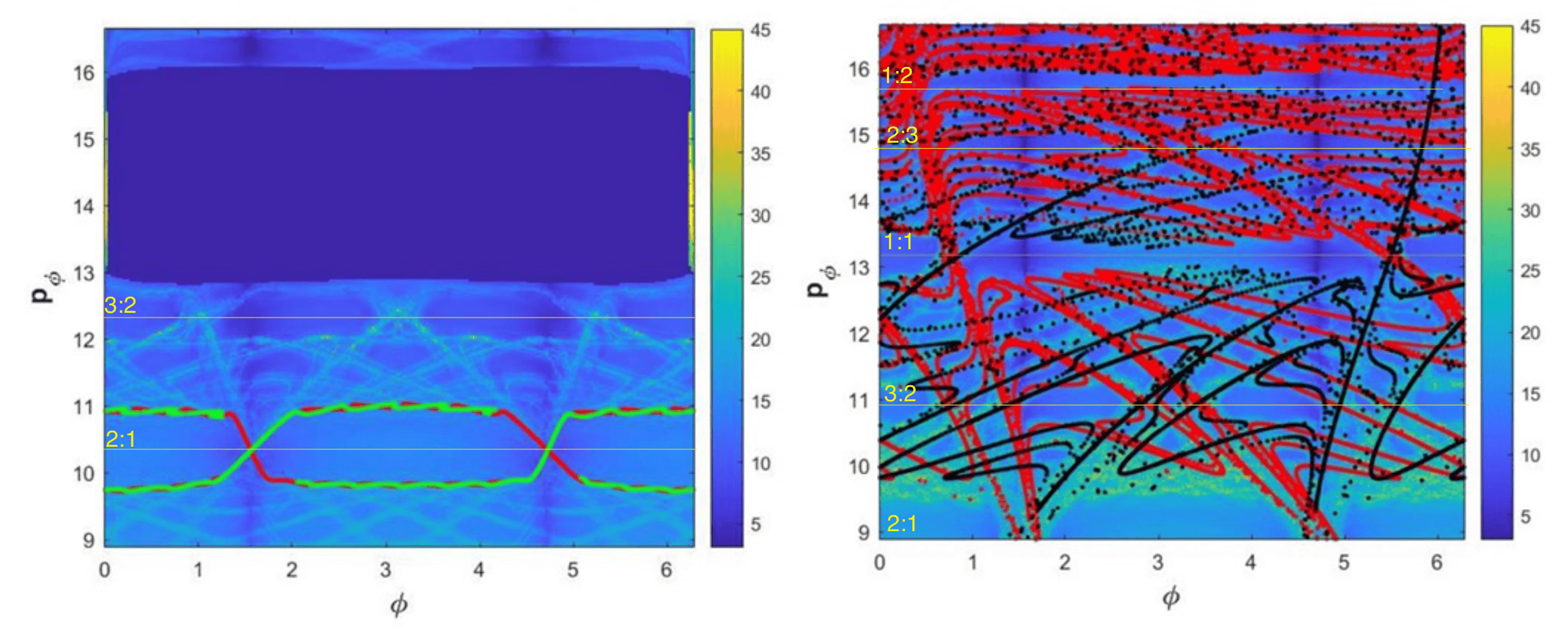}
\caption{\small Overlap of FLI maps with 2:1 resonance stable-unstable manifolds. Left: $T_J = 3.04$ (red-unstable, green-stable); Right: $T_J = 2.84$ (red-unstable, black-stable).The FLI values are shown on a logarithmic scale following the color bar on the right. }
\label{fig:flimanif142152}
\end{figure}

Figure \ref{fig:flimanif} summarizes the main result from the above computation. The stable and unstable manifolds of the unstable periodic orbits of various important MMRs (as indicated in the figure) are shown together with the forward and backward FLI maps (same as in the last panel of Fig.\ref{fig:portraitfli148}) for the Jacobi energy $E_J=-1.48$. These figures show the intricate set of heteroclinic connections between the manifolds of the unstable periodic orbits covering essentially the entire outer asteroidal belt beyond the 2:1 MMR, and reaching all exterior MMRs beyond the co-orbital resonance and up to 2:3. In particular, we have the following properties of the heteroclinic structure:\\
\\
\noindent
(i) The union of all the manifolds (curves $\mathcal{C}^{S}_{q:p,E_J}$, $\mathcal{C}^{U}_{q:p,E_J}$ with $q:p=~$2:1, 3:2, 1:1 \mbox{(orbit PL3)}, 2:3 cover essentially the entire chaotic saddle observed by the numerical FLI maps.\\
\\
\noindent
(ii) The manifolds of the Lyapunov orbit PL3 of the co-orbital (1:1) resonance penetrate deeply in the domains of both the Hilda (3:2) and Hecuba (2:1) interior MMRs, creating heteroclinic channels which essentially allow for the flow of particles throughout all three resonances. In the same way, the manifolds of the PL3 orbit creare heteroclinic connections with the outer MMMs 3:4, 2:3, etc. By the shadowing lemma, close to the doubly asymptotic trajectories associated with these heteroclinic connections, we have chaotic trajectories which undergo `resonance hoping' all the way between the 2:1 and 2:3 resonances, passing through the co-orbital resonance. Physically, these implies trajectories which get temporarily captured in trojan-like tadpole or horseshoe trajectories, originating from one of the interior MMRs (2:1 or 3:2) and tending to an outer resonance (typically 2:3), or vice versa. \\
\\
\noindent
(iii) As shown in the last panel of Fig.\ref{fig:flimanif}, besides the indirect (through the 1:1 resonance) connections, we also have direct heteroclinic connections between the manifolds of the 2:1 with the 3:2 resonance. Physically, this gives the possibility of an immediate transition from an orbit exterior to an orbit interior to the one of Jupiter, without temporary capture at the 1:1 resonance. As shown in Fig.\ref{fig:flimanif142152}, these direct heteroclinic connections become prevalent as the Jacobi energy increases (i.e. as $T_J$ decreases below the limiting value $T_J=3$). On the other hand, for values $E_J<E_{J,L3}$, communication between the interior and exterior MMRs is no longer energetically allowed. In this case the FLI maps still indicate possible connections between the inner MMRs 2:1 and 3:2. However, computing the manifolds of the 2:1 resonance we find that the manifolds' homoclinic lobes are rather constrained around the resonance, thus their heteroclinic connections with other resonances of the outer asteroid belt should affect chaotic trajectories occupying a very small measure within the chaotic saddle. 

\subsection{Manifold reconstruction of the `arches of chaos'}
\label{subsec:arches}

\begin{figure}[h]
\centering
\includegraphics[scale=0.55]{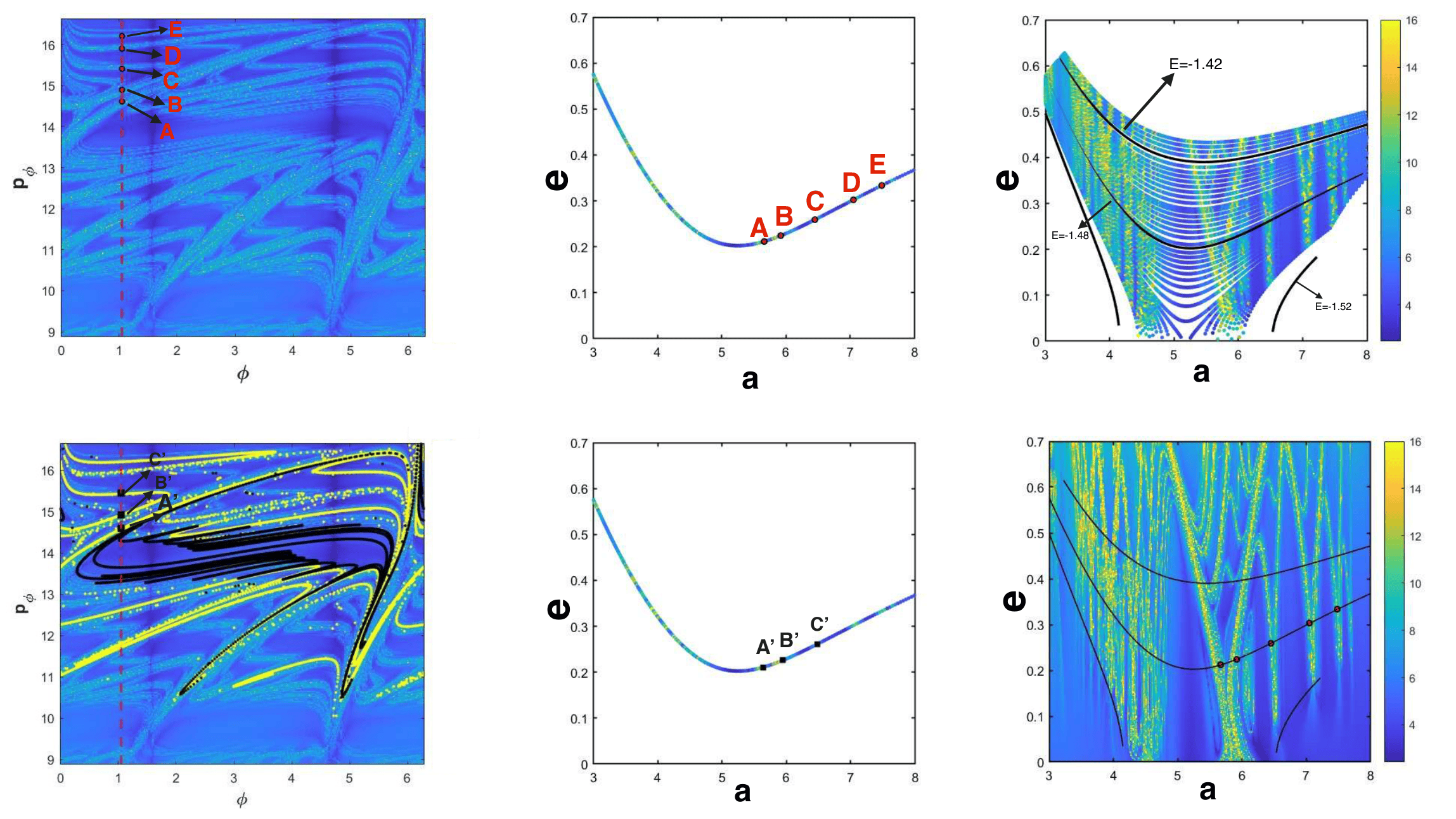}
\caption{\small \textit{Top-left} The FLI map for $E_J=-1.48$ and forward integrated trajectories (same as in the bottom-left panel of Fig.\ref{fig:portraitfli148}). The points A,B,C,D,and E are chosen at local ridges of the FLI map intersecting the vertical line $\varphi=\pi/3$. \textit{Top-middle} The FLI color map along the line $\varphi=\pi/3$ is mapped to a curve the $(a,e)$ representation of the section (see text). \textit{Top-right} Superposition of 40 curves (and the corresponding FLI color maps) computed through the mapping $(\varphi,p_\varphi)\rightarrow(a,e)$ for $\varphi=\pi/3$ and various equidistant values of the Jacobi energy ranging from $E_J=-1.52$ to $E_J=-1.40$. \textit{Bottom-left} The stable manifolds of the 2:3 exterior periodic orbit superposed to the FLI map of the top-left panel. The points $A',B',C'$ are taken on the manifold. \textit{Bottom-middle} same as in the top middle panel, but showing the mapping of the points $A',B',C'$ in the $(a,e)$ plane for the adopted section. \textit{Bottom-right} The entire FLI map for forward integrated trajectories with initial conditions in a grid in the $(a,e)$ plane for the section $\varphi=\pi/3$. The `arches of chaos' are produced by the union of all curves (and their corresponding ridge-located points like A,B,C,D, and E) as those of the top-right panel, but for continous variation of the Jacobi energy $E_J$. }
\label{fig:arches}
\end{figure}
According to \cite{todetal2020}, the `arches of chaos' are manifold structures observed in the projection to the plane $(a,e)$ (semimajor axis - eccentricity) of a particular section of the phase space. Such structures are conveniently illustrated through FLI maps as in figure 1 of \cite{todetal2020}, which shows the FLI maps for forward-integrated trajectories in the plane $(a,e)$ for the section $\ell=\pi/3, \omega=\omega_{Jupiter}$ (which can be set equal to zero without loss of generality). In the present section we reconstruct the `arches of chaos' for a slightly different section, namely $\ell=0$ (pericentric section) and $\omega=\pi/3$. Our aim it to show how the manifolds of the various MMRs under study contribute to the chaotic saddle structures seen in the plane $(a,e)$ for various values of the Jacobi energy $E_J$. 

To this end, consider first the top-left panel of Fig.\ref{fig:arches} (same as the bottom-left panel of Fig.\ref{fig:portraitfli148}), which shows the FLI map in the surface of section for forward-integrated trajectories for $E_J=-1.48$. Consider the vertical slice of initial conditions corresponding to the angle $\varphi=\pi/3$ (red line). The various structures formed by the stable manifolds of several exterior MMRs form `ridges' of the FLI map (see \cite{guzleg2023}) which intersect this slice, for example, at the points marked A,B,C,D,E. Since $E_J>E_{J,L3}$, the entire red line can be mapped to one continuous curve with a global minimum in the plane $(a,e)$, where $a$ is the baricentric semimajor axis of the asteroid. The mapping $(\varphi,p_\varphi)\rightarrow(a,e)$ in the section is computed through the relations $p_\varphi=\sqrt{\mathcal{G}(M_S)(1-e^2)}$, $a=-\mathcal{G}M_s/(2\mathcal{E}_{K})$ where $\mathcal{E}_K=p_{\varphi}^2/(2r^2)-\mathcal{G}M_S/r$ and $r$ is computed through Eq.(\ref{eq:poinc}). The curve is shown in the top-middle panel of Fig.\ref{fig:arches}, with every point on the curve illustrated with the same color-map as for the FLI. The points A,B,C,D,E of the original vertical line are mapped to respective points along the curve. 

The bottom-left panel in Fig.\ref{fig:arches} shows, now, the stable manifolds of the unstable periodic orbit of the 2:3 exterior MMR (yellow) and of the short-period orbit PL3 (black) superposed to the FLI-map of the surface of section at the same value of the Jacobi energy. The manifold traces the whole chaotic saddle structure seen in the FLI map. In particular, the points marked $A',B',C'$ are along the red slice $\varphi=\pi/3$, and they are close to the points A,B,C discussed above, but strictly on the manifold curve ${\cal C}_{2:3,E_J=-1.48}$. The bottom middle panel shows again the slice  $\varphi=\pi/3$ mapped to the plane $(a,e)$, this time with the points $A',B',C'$ marked on it. The main observation is that both the points $A$ to $E$ on the `ridges' of the FLI map, as well as the exaxt points $A',B',C'$ on the manifold curve ${\cal C}_{2:3,E_J=-1.48}$ yield local maxima of the values of the FLI. Along the curve, we have a local maximum of the FLI at any point where the manifold $\mathcal{C}_{2:3,E_J=-1.48}^S$ intersects the slice. Consider, finally, the family of all V-shaped curves in the projection of the section $\varphi=\pi/3$ to the $(a,e)$ plane for various values of $E_J$, as in the top-right panel of Fig.\ref{fig:arches}. Joining the ridge or manifold points of nearby curves allow then to identify the origin of the `arches of chaos', assigning to each `arch' the manifold of the corresponding MMR which locally gives rise to a ridge in the FLI map.    

\subsection{Heteroclinic trajectories and `resonance hopping'}
\label{subsec:reshop}

\begin{figure}[h]
\centering
\includegraphics[scale=0.55]{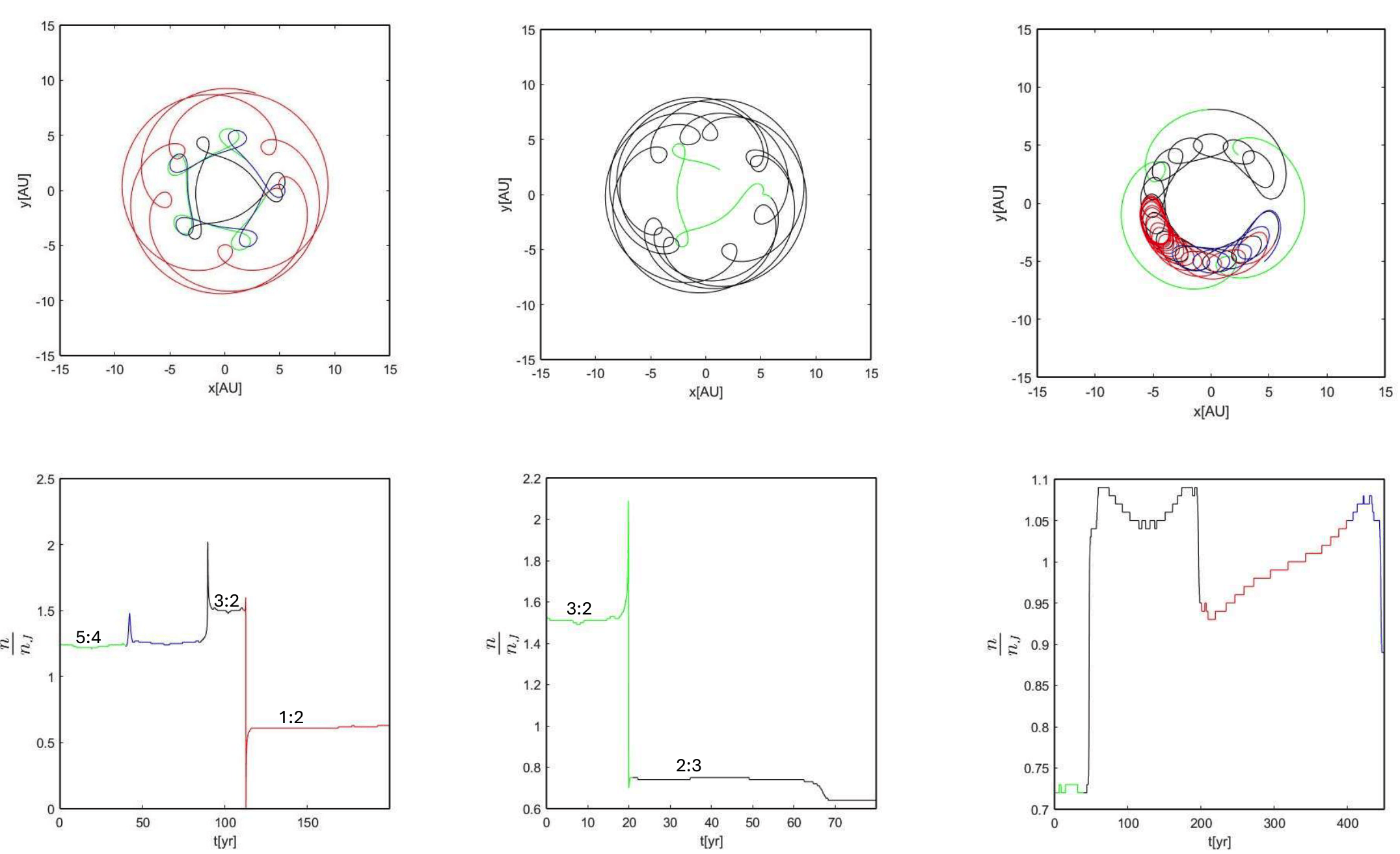}
\caption{\small Three trajectories (top line) selected with initial conditions in the Poincar\'{e} section along the line $\varphi=\pi/3$ and exhibiting `resonance hopping'. The bottom line shows the evolution of the ratio $n/n_J = (a_J/a)^{3/2}$ in the time interval in which the trajectories are integrated. The approximate plateau's in the ratio $n/n_J$ correspond to intervals of permanence of a trajectory in the neighborhood of a particular resonance $p:q$, as indicated in each panel.}
\label{fig:reshop}
\end{figure}
As mentioned above, the intersections between the stable and unstable manifolds of several interior or exterior MMRs indicate the existence of orbits undergoing `resonance hopping' by shadowing the corresponding heteroclinic paths. Figure \ref{fig:reshop} shows example of such heteroclinic connections. A careful investigation of these figures shows that several transitions take place at instances of a close encounter of the trajectories with Jupiter, as, for example, the classical $3:2\rightarrow 2:3$ transition observed in the middle panels of Fig.\ref{fig:reshop}. However, we also find cases where the transition from the exterior to the interior of the orbit of Jupiter takes place without an obvious close encounter with Jupiter. Such is the case, for example, of the transition seen in the right panels of Fig.\ref{fig:reshop}. In this case, coming from the exterior of the co-orbital resonance, the orbit enters first in the horseshoe domain of the co-orbital resonance, and approaches the domain surrounding the Lagrangian point $L5$ through the stable manifold of the unstable periodic orbit $PL3$. At the closest approach to $L3$ the orbit remains temporarily sticky, and then describes half a loop around L5, in the direction from the exterior to the interior of the orbit of Jupiter. 

The behavior exhibited by the trajectories of Fig.\ref{fig:reshop} is typical, and nearly all trajectories with initial conditions along the same section line are observed to undergo some resonant hopping in time intervals ranging between few years and few decades. 

\section{Manifold structure in the ERTBP}
\label{sec:ertbp}
\begin{figure}[h]
\centering
\includegraphics[scale=0.6]{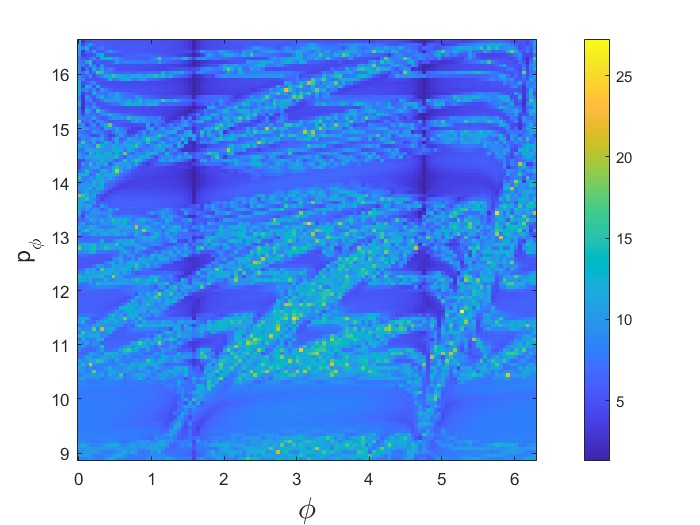}
\caption{\small The FLI map for forward-integrated trajectories taken in a grid of initial conditions exactly equal to those of Fig.\ref{fig:portraitfli148}, but now integrated under the equations of motion of the ERTBP. The manifold structure indicated in this figure is similar to the one of the CRTBP case (bottom-left panel of Fig.\ref{fig:portraitfli148}).}
\label{fig:fliertbp}
\end{figure}
As a final computation, Fig.\ref{fig:fliertbp} shows The FLI map computed for forward-integrated trajectories taken in a grid of initial conditions exactly equal as the one of Fig.\ref{fig:portraitfli148}, integrating the trajectories with the equations of motion of the ERTBP as derived in subsection \ref{subsec:ertbp}. The LCS computed by the union of all ridges of the FLI map in this figure gives a picture very similar to the manifold structure found in the corresponding FLI map under the CRTBP (bottom-left panel of Fig.\ref{fig:portraitfli148}). In particular, we note that most heteroclinic paths such as the paths PL1U, PL1D marked in Fig.\ref{fig:portraitfli148} survive with their structure essentially unaltered when passing from the CRTBP to the ERTBP. 

\section{Conclusions}
\label{sec:conclusions}
In summary, in the present paper we show explicit numerical calculations of the stable and unstable invariant manifolds associated with important first order MMRs as the 2:1 and 3:2, interior, and 2:3, exterior to the orbit of Jupiter. In addition, through short-time FLI computations we obtain a representation of the entire manifold structure (or `Lagrangian Coherent Structure') obtained by these manifolds in the extented resonance overlap fomain transcending the phase space beyond the 2:1 MMR. We give numerical evidence of the fact that the heteroclinic connections between the stable and unstable manifolds of different MMRs may account for the phenomenon of `resonance hopping', conjectured for several real small Solar System bodies. Our main conclusions are the following: 

1. Explicit numerical computations of the stable and unstable manifolds of the first order MMRs $q:p$, $|q-p|=1$ produce curves on a suitable defined (pericentric) surface of section whose union overlaps excellently with the observed LCSs, i.e., the ridges of the FLI maps. These structures visualize immediately and provide insight over the rich network of heteroclinic connections which trascend the entire resonance overlap domain at both the interior and exterior of the orbit of Jupiter at Jacobi energies corresponding to Tisserand parameters $T<3$. Several `heteroclinic paths' allowing direct communication between the exterior ($q<p$) and interior ($q>p$) MMRs are discussed in section \ref{sec:fli}. Such paths are qualitatively similar to the well known heteroclinic connections involving the tube manifolds of the Lyapunov orbits around the collinear Lagrangian points L1 and L2 (as discussed in \cite{koon2000}), \cite{koonetal2001}. However, the connections found here allow in principle for `resonance hopping' that does not involve the above tube manifolds, but only the manifolds of the associated outgoing and incoming MMRs. 

2. We explain how the manifold structures and the LCSs observed in our FLI maps allow to interpret the `arches of chaos' (\cite{todetal2020}), which essentially correspond to the same structures viewed in a suitable section and visualized in the plane of asteroid orbital elements $(a,e)$. 

3. Our main core of results is within the framework of the planar and circular RTBP. However, we give numerical evidence that the manifold structures observed in the circular problem persist in an essentially unaltered form when transitioning from the circular to the elliptic RTBP. We also indicate a possible applicability of our results to families of Solar System small bodies transiting between MMRs in the outer asteroid belt or through the orbit of Jupiter, as are the quasi-Hildas, JFCs and Centaurs. However, caution should be payed on firmly assuming such a connection, since our study is limited to the planar RTBP, while several of the above bodies have orbits substantially inclined with respect to the Sun-Jupiter orbital plane, and also influenced by the direct or indirect perturbing effects of the remaining giant planets in the solar system. In fact, the persistence against perturbations of the manifold structures here observed are a good hint towards a wider applicability of results as the ones here reported. This subject is a natural next step for further study. \\
\\
\noindent
{\bf Acknowledgements} C.E. acknowledges several inspiring discussions with A. Rosengren, N. Todorovic and M. Guzzo. A.F.G. was supported by the Italian national inter-university
PhD programme in Space Science and Technology and INdAM group G.N.F.M.

\bibliographystyle{plain}
\bibliography{bibliografia}

@ARTICLE{baimal2009,
       author = {{Bailey}, B. L. and {Malhotra}, R.},
        title = "{Two dynamical classes of Centaurs}",
      journal = {Icarus},
         year = 2009,
       
       volume = {203},
       number = {1},
        pages = {155-163},
          doi = {10.1016/j.icarus.2009.03.044},
archivePrefix = {arXiv},
       eprint = {0906.4795},
 primaryClass = {astro-ph.EP},
       adsurl = {https://ui.adsabs.harvard.edu/abs/2009Icar..203..155B}
}

@ARTICLE{belmar1997,
       author = {{Belbruno}, E. and {Marsden}, B. G.},
        title = "{Resonance Hopping in Comets}",
      journal = {Astronomical Journal},
         year = 1997,
        
       volume = {113},
        pages = {1433},
          doi = {10.1086/118359},
       adsurl = {https://ui.adsabs.harvard.edu/abs/1997AJ....113.1433B}
}

@ARTICLE{beletal2008,
       author = {{Belbruno}, E. and {Topputo}, F. and {Gidea}, M.},
        title = {Resonance transitions associated to weak capture in the restricted three-body problem},
      journal = {Advances in Space Research},
         year = {2008},
        
       volume = {42},
       number = {8},
        pages = {1330-1351},
          doi = {10.1016/j.asr.2008.01.018},
       adsurl = {https://ui.adsabs.harvard.edu/abs/2008AdSpR..42.1330B}
}

@ARTICLE{cavallariefhty,
       author = {{Cavallari}, I. and {Efthymiopoulos}, C.},
        title = "{Closed-form perturbation theory in the restricted three-body problem without relegation}",
      journal = {Celestial Mechanics and Dynamical Astronomy},
         year = 2022,
       volume = {134},
       number = {2},
          eid = {16},
        pages = {16},
          doi = {10.1007/s10569-022-10070-y},
archivePrefix = {arXiv},
       eprint = {2110.14489},
 primaryClass = {astro-ph.EP},
       adsurl = {https://ui.adsabs.harvard.edu/abs/2022CeMDA.134...16C}
}

@article{chaetal2022,
       author = {{Chandler}, C. and {Oldroyd}, W. and {Trujillo}, C.},
        title = "{282P/323137: A Migrating and Active Quasi-Hilda Object}",
    booktitle = {American Astronomical Society Meeting Abstracts \#241},
         year = 2023,
       journal = {American Astronomical Society Meeting Abstracts},
       volume = {241},
       
          eid = {136.04},
        pages = {136.04},
       adsurl = {https://ui.adsabs.harvard.edu/abs/2023AAS...24113604C}
}

@ARTICLE{chi1979,
       author = {{Chirikov}, B. V.},
        title = "{A universal instability of many-dimensional oscillator systems}",
      journal = {Physics Reports},
         year = 1979,
        
       volume = {52},
       number = {5},
        pages = {263-379},
          doi = {10.1016/0370-1573(79)90023-1},
       adsurl = {https://ui.adsabs.harvard.edu/abs/1979PhR....52..263C}
}

@article{coretal2024,
author = {Correa-Otto, J. and García-Migani, E. and Gil-Hutton, R.},
year = {2023},

pages = {},
title = {The population of Comet candidates among quasi-Hilda objects revisited and updated},
volume = {527},
journal = {Monthly Notices of the Royal Astronomical Society},
doi = {10.1093/mnras/stad3234}
}

@article{decketal2013,
doi = {10.1088/0004-637X/774/2/129},
url = {https://doi.org/10.1088/0004-637X/774/2/129},
year = {2013},

publisher = {The American Astronomical Society},
volume = {774},
number = {2},
pages = {129},
author = {Deck, K. M. and Payne, M. and Holman, M. J.},
title = {FIRST-ORDER RESONANCE OVERLAP AND THE STABILITY OF CLOSE TWO-PLANET SYSTEMS},
journal = {The Astrophysical Journal},
}

@ARTICLE{deletal2005,
author = {Dellnitz, M. and Junge, O. and Lo, M. and Marsden, J. and Padberg-Gehle, K. and Preis, R. and Ross, S. and Thiere, B.},
year = {2005},


title = {Transport of Mars-Crossing Asteroids from the Quasi-Hilda Region},
volume = {94},
journal = {Physical review letters},
doi = {10.1103/PhysRevLett.94.231102}
}

@ARTICLE{disisetal2005,
       author = {{Di Sisto}, R. P. and {Brunini}, A. and {Dirani}, L. D. and {Orellana}, R. B.},
        title = "{Hilda asteroids among Jupiter family comets}",
      journal = {Icarus},
         year = 2005,
        
       volume = {174},
       number = {1},
        pages = {81-89},
          doi = {10.1016/j.icarus.2004.10.024},
       adsurl = {https://ui.adsabs.harvard.edu/abs/2005Icar..174...81D}
}

@ARTICLE{dunqui1993,
       author = {{Duncan}, M. J. and {Quinn}, T.},
        title = "{The long-term dynamical evolution of the solar system.}",
      journal = {Annual Review of Astronomy and Astrophysics},
         year = 1993,
        
       volume = {31},
        pages = {265-295},
          doi = {10.1146/annurev.aa.31.090193.001405},
       adsurl = {https://ui.adsabs.harvard.edu/abs/1993ARA&A..31..265D}
}

@ARTICLE{dun1989,
       author = {{Duncan}, M. and {Quinn}, T. and {Tremaine}, S.},
        title = "{The long-term evolution of orbits in the solar system: A mapping approach}",
      journal = {Icarus},
         year = 1989,
        
       volume = {82},
       number = {2},
        pages = {402-418},
          doi = {10.1016/0019-1035(89)90047-X},
       adsurl = {https://ui.adsabs.harvard.edu/abs/1989Icar...82..402D}
}

@ARTICLE{dvokub2022,
       author = {{Dvorak}, R. and {Kubala}, M.},
        title = "{Are there long-living Hilda-like asteroids between Jupiter and Saturn?}",
      journal = {Astronomische Nachrichten},
         year = 2022,
        
       volume = {343},
         
        pages = {20009},
          doi = {10.1002/asna.20220009},
       adsurl = {https://ui.adsabs.harvard.edu/abs/2022AN....34320009D}
}

@ARTICLE{feretal1996,
       author = {{Ferraz-Mello}, S. and {Klafke}, J.~C. and {Michtchenko}, T.~A. and {Nesvorn{\'y}}, D.},
        title = "{Chaotic Transitions in Resonant Asteroidal Dynamics}",
      journal = {Celestial Mechanics and Dynamical Astronomy},
         year = 1996,
        
       volume = {64},
       number = {1-2},
        pages = {93-105},
          doi = {10.1007/BF00051608},
       adsurl = {https://ui.adsabs.harvard.edu/abs/1996CeMDA..64...93F}
}

@book{FerrazMello2024,
  author    = {{Ferraz-Mello}, S.},
  title     = {Chaotic Dynamics in Planetary Systems},
  series    = {Springer Praxis Books},
  publisher = {Springer Cham},
  year      = {2024},
  isbn      = {978-3-031-45815-6},
  doi       = {10.1007/978-3-031-45816-3},
  edition   = {1},
  pages     = {167},
  note      = {XI, 167 pages}
}

@ARTICLE{fitros2022,
author = {Fitzgerald, J. and Ross, S.},
year = {2022},

pages = {},
title = {Geometry of transit orbits in the periodically-perturbed restricted three-body problem},
volume = {70},
journal = {Advances in Space Research},
doi = {10.1016/j.asr.2022.04.029}
}

@article{fraetal1993,
       author = {{Franklin}, F. and {Lecar}, M. and {Murison}, M.},
        title = "{Chaotic Orbits and Long Term Stability: an Example From Asteroids of the Hilda Group}",
      journal = {Astronomical Journal},
         year = 1993,
       
       volume = {105},
        pages = {2336},
          doi = {10.1086/116611},
       adsurl = {https://ui.adsabs.harvard.edu/abs/1993AJ....105.2336F}
}

@ARTICLE{fra1995,
       author = {{Franklin}, F.},
        title = "{Chaotic Orbits Within the 3-2 Jovian Mean Motion Resonance}",
      journal = {Astronomical Journal},
         year = {1995},
        
       volume = {110},
        pages = {1879},
          doi = {10.1086/117659},
       adsurl = {https://ui.adsabs.harvard.edu/abs/1995AJ....110.1879F}
}

@article{frosch1979,
       author = {{Froeschl\'e}, C. and {Scholl}, H.},
        title = "{New Numerical Experiments to Deplete the Outer Part of the Asteroidal Belt}",
      journal = {Astronomy and Astrophysics},
         year = 1979,
        
       volume = {72},
       number = {1-2},
        pages = {246-255},
       adsurl = {https://ui.adsabs.harvard.edu/abs/1979A&A....72..246F}
}

@article{froschetal2000,
       author = {{Froeschl\'e}, C. and {Guzzo}, M. and {Lega}, E.},
        title = "{Graphical evolution of the Arnold web: from order to chaos}",
      journal = {Science},
         year = 2000,
       volume = {289},
        pages = {2108-2110}
}

@article{gargil2018,
       author = {{Garc{\'\i}a-Migani}, E. and {Gil-Hutton}, R.},
        title = "{The activity and dynamical evolution of quasi-hilda asteroid (457175) 2008 GO98}",
      journal = {Planetary and Space Science},
         year = 2018,
        
       volume = {160},
        pages = {12-18},
          doi = {10.1016/j.pss.2018.03.011},
       adsurl = {https://ui.adsabs.harvard.edu/abs/2018P&SS..160...12G}
}

@article{gawmar2009,
author = {Gawlik, E. and Marsden, J. and Du Toit, P. and Campagnola, S.},
year = {2009},

pages = {227-249},
title = {Lagrangian coherent structures in the planar elliptic restricted three-body problem},
volume = {103},
journal = {Celestial Mechanics and Dynamical Astronomy},
doi = {10.1007/s10569-008-9180-3}
}

@ARTICLE{gupkum2016,
       author = {{Gupta}, B. R. and {Kumar}, V.},
        title = "{Characterization of the Phase Space Structure of Circular Restricted Three-Body Problem: An Alternative Approach}",
      journal = {International Journal of Bifurcation and Chaos},
         year = 2016,
        
       volume = {26},
       number = {2},
          eid = {1650029-1351},
        pages = {1650029-1351},
          doi = {10.1142/S0218127416500292},
       adsurl = {https://ui.adsabs.harvard.edu/abs/2016IJBC...2650029G}
}

@article{guzleg2014,
  author = {{Guzzo}, M. and {Lega}, E.},
  title={Evolution of the Tangent Vectors and Localization of the Stable and Unstable Manifolds of Hyperbolic Orbits by Fast Lyapunov Indicators},
  journal={SIAM J. Appl. Math.},
  year={2013},
  volume={74},
  pages={1058-1086},
  url={https://api.semanticscholar.org/CorpusID:42132244}
}

@ARTICLE{guzleg2023,
       author = {{Guzzo}, M. and {Lega}, E.},
        title = "{Theory and applications of fast Lyapunov indicators to model problems of celestial mechanics}",
      journal = {Celestial Mechanics and Dynamical Astronomy},
         year = 2023,
        
       volume = {135},
       number = {4},
          eid = {37},
        pages = {37},
          doi = {10.1007/s10569-023-10152-5},
       adsurl = {https://ui.adsabs.harvard.edu/abs/2023CeMDA.135...37G}
}

@article{hal2002,
title = {Lagrangian coherent structures and mixing in two-dimensional turbulence},
journal = {Physica D: Nonlinear Phenomena},
volume = {147},
number = {3},
pages = {352-370},
year = {2000},
issn = {0167-2789},
doi = {https://doi.org/10.1016/S0167-2789(00)00142-1},
url = {https://www.sciencedirect.com/science/article/pii/S0167278900001421},
author = {G. Haller and G. Yuan},
}

@article{jornic2020,
author = {Jorba, A. and N., Begoña},
year = {2020},

pages = {105327},
title = {Transport and invariant manifolds near L3 in the Earth-Moon Bicircular model},
volume = {89},
journal = {Communications in Nonlinear Science and Numerical Simulation},
doi = {10.1016/j.cnsns.2020.105327}
}

@article{kazkaz2021,
       author = {{Kazantsev}, A. and {Kazantseva}, L-},
        title = "{On a possibility of transfer of asteroids from the 2:1 mean motion resonance with Jupiter to the Centaur zone}",
      journal = {Monthly Notices of the Royal Astronomical Society},
         year = 2021,
        
       volume = {505},
       number = {1},
        pages = {408-414},
          doi = {10.1093/mnras/stab1078},
archivePrefix = {arXiv},
       eprint = {2105.14062},
 primaryClass = {astro-ph.EP},
       adsurl = {https://ui.adsabs.harvard.edu/abs/2021MNRAS.505..408K}
}

@article{koon2000,
author = {Koon, W. and Lo, M.W. and Marsden, J.E. and Ross, S.D.},
year = {2000},

pages = {427-469},
title = {Heteroclinic Connections Between Periodic Orbits and Resonance Transitions in Celestial Mechanics},
volume = {10},
journal = {Chaos},
doi = {10.1063/1.166509}
}

@article{koonetal2001,
       author = {{Koon}, W. and {Lo}, M.W. and {Marsden}, J.E. and {Ross}, S.D.},
        title = "{Resonance and Capture of Jupiter Comets}",
      journal = {Celestial Mechanics and Dynamical Astronomy},
         year = 2001,
        
       volume = {81},
        pages = {27-38},
          doi = {10.1023/A:1013398801813},
       adsurl = {https://ui.adsabs.harvard.edu/abs/2001CeMDA..81...27K}
}

@article{kumetal2024,
        title = {Investigation of Interior Mean Motion Resonances and Heteroclinic Connections in the Earth-Moon System},
        year = {2024},
       author = {Kumar, B. and Rawat, A. and Rosengren, A. and Ross, S.},
}

@article{lecetal1992,
       author = {{Lecar}, M. and {Franklin}, F. and {Murison}, M.},
        title = "{On Predicting Long-Term Orbital Instability: A Relation Between the Lyapunov Time and Sudden Orbital Transitions}",
      journal = {Astronomical Journal },
         year = 1992,
        
       volume = {104},
        pages = {1230},
          doi = {10.1086/116312},
       adsurl = {https://ui.adsabs.harvard.edu/abs/1992AJ....104.1230L}
}

@ARTICLE{leixu2018,
       author = {{Lei}, H. and {Xu}, B.},
        title = "{Resonance transition periodic orbits in the circular restricted three-body problem}",
      journal = {Astrophysics and Space Science},
         year = 2018,
        
       volume = {363},
       number = {4},
          eid = {70},
        pages = {70},
          doi = {10.1007/s10509-018-3290-5},
       adsurl = {https://ui.adsabs.harvard.edu/abs/2018Ap&SS.363...70L}
}

@ARTICLE{michfer1995,
       author = {{Michtchenko}, T.~A. and {Ferraz-Mello}, S.},
        title = "{Comparative study of the asteroidal motion in the 3:2 and 2:1 resonances with Jupiter. I. Planar model.}",
      journal = {Astronomy and Astrophysics},
         year = 1995,
        
       volume = {303},
        pages = {945},
       adsurl = {https://ui.adsabs.harvard.edu/abs/1995A&A...303..945M}
}

@ARTICLE{milnob1993,
       author = {{Milani}, A. and {Nobili}, A. M.},
        title = "{Asteroid 522 Helga is Chaotic and Stable}",
      journal = {Celestial Mechanics and Dynamical Astronomy},
         year = 1993,
       
       volume = {56},
       number = {1-2},
        pages = {323-324},
          doi = {10.1007/BF00699743},
       adsurl = {https://ui.adsabs.harvard.edu/abs/1993CeMDA..56..323M}
}

@ARTICLE{miletal1997,
       author = {{Milani}, A. and {Nobili}, A.M. and {Kne{\v{z}}evi{\'c}}, Z.},
        title = "{Stable Chaos in the Asteroid Belt}",
      journal = {Icarus},
         year = 1997,
       volume = {125},
        pages = {13-31}
}

@BOOK{murder1999,
       author = {{Murray}, C. D. and {Dermott}, S. F.},
        title = "{Solar System Dynamics}",
         year = 1999,
          doi = {10.1017/CBO9781139174817},
       adsurl = {https://ui.adsabs.harvard.edu/abs/1999ssd..book.....M}
}

@ARTICLE{muswya2012,
       author = {{Mustill}, A. J. and {Wyatt}, M. C.},
        title = "{Dependence of a planet's chaotic zone on particle eccentricity: the shape of debris disc inner edges}",
      journal = {Monthly Notices of the Royal Astronomical Societ},
         year = 2012,
       
       volume = {419},
       number = {4},
        pages = {3074-3080},
          doi = {10.1111/j.1365-2966.2011.19948.x},
archivePrefix = {arXiv},
       eprint = {1110.1282},
 primaryClass = {astro-ph.EP},
       adsurl = {https://ui.adsabs.harvard.edu/abs/2012MNRAS.419.3074M}
}

@ARTICLE{nesfer1997,
       author = {{Nesvorn{\'y}}, D. and {Ferraz-Mello}, S.},
        title = "{On the Asteroidal Population of the First-Order Jovian Resonances}",
      journal = {Icarus},
         year = 1997,
        
       volume = {130},
       number = {2},
        pages = {247-258},
          doi = {10.1006/icar.1997.5807},
       adsurl = {https://ui.adsabs.harvard.edu/abs/1997Icar..130..247N}
}

@ARTICLE{oldetal2023,
       author = {{Oldroyd}, W. J. and {Chandler}, C. O. and {Trujillo}, C. A. and {Sheppard}, S. S. and {Hsieh}, H. H. and {Kueny}, J. K. and {Burris}, W. A. and {DeSpain}, J. A. and {Farrell}, K. A. and {Mazzucato}, M. T. and {Bosch}, Milton K.~D. and {Shaw-Diaz}, T. and {Gonano}, V.},
        title = "{Recurring Activity Discovered on Quasi-Hilda 2009 DQ118}",
      journal = {Astrophysical Journal Letters},
         year = 2023,
        
       volume = {957},
       number = {1},
          
          doi = {10.3847/2041-8213/acfcbc},
archivePrefix = {arXiv},
       eprint = {2311.02160},
 primaryClass = {astro-ph.EP},
       adsurl = {https://ui.adsabs.harvard.edu/abs/2023ApJ...957L...1O}
}

@ARTICLE{paeeft2015,
       author = {{P{\'a}ez}, R. and {Efthymiopoulos}, C.},
        title = "{Trojan resonant dynamics, stability, and chaotic diffusion, for parameters relevant to exoplanetary systems}",
      journal = {Celestial Mechanics and Dynamical Astronomy},
         year = 2015,
        
       volume = {121},
       number = {2},
        pages = {139-170},
          doi = {10.1007/s10569-014-9591-2},
archivePrefix = {arXiv},
       eprint = {1410.1407},
 primaryClass = {astro-ph.EP},
       adsurl = {https://ui.adsabs.harvard.edu/abs/2015CeMDA.121..139P}
}

@article{paeguz2023,
author = {{P{\'a}ez}, R. and Guzzo, M.},
year = {2022},

pages = {133402},
title = {On the semi-analytical construction of halo orbits and halo tubes in the elliptic restricted three-body problem},
volume = {439},
journal = {Physica D: Nonlinear Phenomena},
doi = {10.1016/j.physd.2022.133402}
}

@article{panhou2022,
author = {Pan, S. and Hou, X.},
year = {2022},

pages = {},
title = {Analysis of Resonance Transition Periodic Orbits in the Circular Restricted Three-Body Problem},
volume = {12},
journal = {Applied Sciences},
doi = {10.3390/app12188952}
}

@article{quifab2006,
author = {Quillen, A. and Faber, P.},
year = {2006},

pages = {},
title = {Chaotic zone boundary for low free eccentricity particles near an eccentric planet},
volume = {373},
journal = {Monthly Notices of the Royal Astronomical Society},
doi = {10.1111/j.1365-2966.2006.11122.x}
}

@ARTICLE{rametal2015,
       author = {{Ramos}, X.S. and {Correa-Otto}, J.A. and {Beaug{\'e}}, C.},
        title = "{The resonance overlap and Hill stability criteria revisited}",
      journal = {Celestial Mechanics and Dynamical Astronomy},
         year = 2015,
        
       volume = {123},
       number = {4},
        pages = {453-479},
          doi = {10.1007/s10569-015-9646-z},
archivePrefix = {arXiv},
       eprint = {1509.03607},
 primaryClass = {astro-ph.EP},
       adsurl = {https://ui.adsabs.harvard.edu/abs/2015CeMDA.123..453R}
}

@ARTICLE{ricketal2017,
       author = {{Rickman}, H. and {Gabryszewski}, R. and {Wajer}, P. and {Wi\'sniowski}, T. and {W\'ojcikowski}, T. and {Szutowicz}, S. and {Valsecchi}, G.B. and {Morbidelli}, A.},
        title = "{Secular orbital evolution of Jupiter family comets}",
      journal = {Astronomy and Astrophysics},
         year = 2017,
       volume = {598},
        pages = {1-15}
}

@ARTICLE{roigetal2002,
       author = {{Roig}, F. and {Nesvorn{\'y}}, D. and {Ferraz-Mello}, S.},
        title = "{Asteroids in the 2 : 1 resonance with Jupiter: dynamics and size distribution}",
      journal = {Monthly Notices of the Royal Astronomical Society},
         year = 2002,
        
       volume = {335},
       number = {2},
        pages = {417-431},
          doi = {10.1046/j.1365-8711.2002.05635.x},
       adsurl = {https://ui.adsabs.harvard.edu/abs/2002MNRAS.335..417R}
}

@article{rossiefthy,
author = {Rossi, M. and Efthymiopoulos, C.},
year = {2023},

pages = {},
title = {Relegation-free closed-form perturbation theory and the domain of secular motions in the restricted three-body problem},
volume = {135},
journal = {Celestial Mechanics and Dynamical Astronomy},
doi = {10.1007/s10569-023-10154-3}
}

@ARTICLE{saridetal2019,
author = {Sarid, G. and Volk, K. and Steckloff, J. and Harris, W. and Womack, M. and Woodney, L.},
year = {2019},


title = {29P/Schwassmann–Wachmann 1, A Centaur in the Gateway to the Jupiter-family Comets},
volume = {883},
journal = {The Astrophysical Journal Letters},
doi = {10.3847/2041-8213/ab3fb3}
}

@ARTICLE{schu2009,
        title = {Numerical studies of chaotic Hilda-type orbits},
        year = {2009},
       author = {Schubart, J.},
      journal = {Celestial Mechanics and Dynamical Astronomy},
}

@ARTICLE{tanetal1990,
       author = {{Tancredi}, G. and {Lindgren}, M. and {Rickman}, H.},
        title = "{Temporary satellite capture and orbital evolution of Comet P/Helin-Roman-Crockett}",
      journal = {Astronomy and Astrophysics},
         year = 1990,
        
       volume = {239},
       number = {1-2},
        pages = {375-380},
       adsurl = {https://ui.adsabs.harvard.edu/abs/1990A&A...239..375T}
}

@ARTICLE{todetal2020,
author = {Todorovic, N. and Wu, D. and Rosengren, A.},
year = {2020},


title = {The arches of chaos in the Solar System},
volume = {6},
journal = {Science Advances},
doi = {10.1126/sciadv.abd1313}
}

@article{top2008,
       author = {{Topputo}, F. and {Belbruno}, E. and {Gidea}, M.},
        title = "{Resonant motion, ballistic escape, and their applications in astrodynamics}",
      journal = {Advances in Space Research},
         year = 2008,
        
       volume = {42},
       number = {8},
        pages = {1318-1329},
          doi = {10.1016/j.asr.2008.01.017},
       adsurl = {https://ui.adsabs.harvard.edu/abs/2008AdSpR..42.1318T}
}

@article{toth2006,
       author = {{Toth}, I.},
        title = "{The quasi-Hilda subgroup of ecliptic comets – an update}",
      journal = {Astronomy and Astrophysics},
         year = 2006,
       volume = {448},
        pages = {1191-1196}
}

@ARTICLE{ursgal2013,
  title={Instabilities in the Sun–Jupiter–Asteroid three body problem},
  author={J. C. Urschel and J. R. Galante},
  journal={Celestial Mechanics and Dynamical Astronomy},
  year={2013},
  volume={115},
  pages={233-259},
  url={https://api.semanticscholar.org/CorpusID:379055}
}

@article{wanmal2017,
       author = {{Wang}, X. and {Malhotra}, R.},
        title = "{Mean Motion Resonances at High Eccentricities: The 2:1 and the 3:2 Interior Resonances}",
      journal = {Astronomical Journal },
         year = 2017,
        
       volume = {154},
       number = {1},
          eid = {20},
        pages = {20},
          doi = {10.3847/1538-3881/aa762b},
archivePrefix = {arXiv},
       eprint = {1702.02137},
 primaryClass = {astro-ph.EP},
       adsurl = {https://ui.adsabs.harvard.edu/abs/2017AJ....154...20W}
}

@article{wis1980,
       author = {{Wisdom}, J.},
        title = "{The resonance overlap criterion and the onset of stochastic behavior in the restricted three-body problem}",
      journal = {Astronomical Journal},
         year = 1980,
        
       volume = {85},
        pages = {1122-1133},
          doi = {10.1086/112778},
       adsurl = {https://ui.adsabs.harvard.edu/abs/1980AJ.....85.1122W}
}

\end{document}